\documentclass[10pt,twocolumn,a4paper]{aastex63}
\usepackage{amsmath}
\usepackage{natbib}

\def\beq{\begin{eqnarray}}
\def\eeq{\end{eqnarray}}

\received{}
\revised{}
\accepted{}
\submitjournal{ApJ}
\shorttitle{Gravitational waves from hyperaccretion in collapsars}
\shortauthors{Wei \& Liu}
	
\begin{document}

\title{Black hole hyperaccretion in collapsars. II. Gravitational waves}

\correspondingauthor{Tong Liu}
\email{tongliu@xmu.edu.cn}

\author{Yun-Feng Wei}
\affiliation{Department of Astronomy, Xiamen University, Xiamen, Fujian 361005, China}

\author[0000-0001-8678-6291]{Tong Liu}
\affiliation{Department of Astronomy, Xiamen University, Xiamen, Fujian 361005, China}

\begin{abstract}
As progenitors of gamma-ray bursts (GRBs), core collapse of massive stars and coalescence of compact object binaries are believed to be powerful sources of  gravitational waves (GWs). In the collapsar scenario, a rotating stellar-mass black hole (BH) surrounded by a hyperaccretion disk might be running in the center of a massive collapsar, which is one of the plausible central engines of long GRBs. Such a BH hyperaccretion disk would be in a state of a neutrino-dominated accretion flow (NDAF) at the initial stage of the accretion process; meanwhile, the jets attempt to break out from the envelope and circumstellar medium to power GRBs. In addition to collapsars, the BH hyperaccretion systems are important sources of neutrinos and GWs. In this paper, we investigated the GW emission generated by the anisotropic neutrino emission from NDAFs in the collapsar scenarios. As the results indicate, the typical frequency of GWs is $\sim$ 1-100 Hz, and the masses and metallicities of the progenitor stars have slight effects on the GW strains. The GWs from NDAFs might be detected by operational or planned detectors at the distance of 10 kpc. Moreover, comparisons of the detectable GWs from collapsars, NDAFs, and GRB jets (internal shocks) are displayed. By combining the electromagnetic counterparts, neutrinos, and GWs, one may constrain the characteristics of collapsars and central BH accretion systems.
\end{abstract}

\keywords{accretion, accretion disks - black hole physics - gamma-ray burst: general - gravitational waves - neutrinos - star: massive}

\section{Introduction}

The Laser Interferometer Gravitational-Wave Observatory (LIGO) detection of gravitational waves (GWs) from the binary black hole (BH) merger GW150914 marked that we entered an era of GW astronomy \citep{Abbott2016}. The detection of GW event from a binary neutron star (NS) merger system GW170817 \citep{Abbott2016} that was associated with electromagnetic signals marked the beginning of multi-messenger astronomy with GWs. In the future, astrophysical source including massive star collapse, rapidly rotating NSs, and other violent events in the Universe might be detected by GW detectors \citep[e.g.,][]{Cutler2002}. Especially for massive star collapse, such an event is a promising multi-messenger transient source.

Observation evidences have indicated that core collapse massive stars are the progenitors of long-duration gamma-ray bursts \citep[LGRBs, see the review by][]{Woosley2006,Kumar2015}. The majority of LGRB host galaxies are irregular, star-forming galaxies \citep[e.g.,][]{Fruchter2006}. A handful of LGRBs are associated with core collapse supernovae \citep[CCSNe, see, e.g.,][]{Galama1998,Hjorth2003,Stanek2003,Malesani2004,Modjaz2006,Pian2006}. Note that CCSNe are diverse, broadly partitioned in normal  (narrow line) and relatively more energetic (broad line) events \citep[e.g.,][]{Maurer2010,van Putten2011}. The observations show that some broad-lined and bright type Ib/c SNe are accompanied by LGRBs. In the collapsar model \citep[e.g.,][]{Woosley1993,MacFadyen1999,Woosley2002,Zhang2004,Woosley2012}, the core of the massive star will collapse; then a few $M_{\odot}$ BH surrounded by a temporary disk with a very high accretion rate might be formed. As a plausible central engine of LGRBs, this accretion process can launch powerful jets. If the activity of the central engine lasts long enough to allow the jets to break out from the envelope and circumstellar medium, an energetic LGRB will be triggered.

Generally, a BH hyperaccretion system can launch GRB jets via two well-known mechanisms: the neutrino-antineutrino annihilation process and the Blandford-Znajek \citep[BZ,][]{Blandford1977} mechanism. If the accretion rate is very high $(\sim 0.001-10\,M_{\odot }\,\rm{s}^{-1})$, then the photons are trapped in the disk, and generous neutrinos are produced. Neutrino pairs are emitted from the disk surface and annihilate above the disk to power gamma-ray bursts (GRBs). Such an accretion disk is called a neutrino-dominated accretion flow (NDAF), whose properties have been widely investigated over the past decades \citep[e.g.,][]{Popham1999,Narayan2001,Kohri2002,Lee2005,Gu2006,Chen2007,Janiuk2007,Kawanaka2007,Liu2007,Liu2015,Liu2017,Lei2009,Xue2013,Song2016,Nagataki2018}. In the BZ mechanism, the magnetic lines tied on the disk will fall into the BH, followed by the accretion materials; then, Poynting jets would be launched via extraction of the spin energy of the BH to power GRBs \citep[e.g.,][]{Lee2000a,Lee2000b,Mizuno2004,McKinney2004,Barkov2008,Nagataki2009,Lei2013,Lei2017,Wu2013}.

Nevertheless, for the very rapidly rotating BHs surrounded by the magnetized disks or toruses, the temporal evolution of the accretion onto the BHs may subject to the large-scale magnetic torques \citep{van Putten Ostriker2001}. Thus the accretion mode will be changed. The additional spin-up torque from the BH may arrest the inflow for the duration of the BH spin-down lifetime. Therefore, the activity duration of the central engine can derive from both the timescale of accretion flow and the lifetime of the BH spin, which are well-known also in the active galactic nuclei community \citep{O'Dea2002}. \citet{van Putten2001} suggested that LGRBs arise with the rapidly spinning BHs in the suspended accretion, while short GRBs arise with the slowly spinning BHs.

Multimessenger observations are essential to constrain the characteristics of collapsars, especially for the central BH accretion systems. It is difficult to extract the information of the central engine from electromagnetic signals, as the most observed electromagnetic signals from GRBs are produced in the regions far from the central engines \citep{Cutler2002}. Neutrinos and gravitational waves (GWs) can provide us with the information hidden deep inside the stellar cores. The detectable MeV neutrinos from NDAFs have been discussed in \citet{Liu2016}. These neutrinos can reach a luminosity of $10^{50}-10^{51}\,\rm{erg}\,\rm{s}^{-1}$, peaking at $\sim 10$ MeV, and might be observed by the next generation MeV neutrino detectors, such as Hyper-Kamiokande, when the events are close enough to Earth. GW emission from GRB central engines has been investigated in many previous works. \citet{Sun2012} studied the GWs from jet precession driven by an NDAF around a spinning BH. GWs generated by the anisotropic neutrino emission from NDAFs have been discussed in some studies \citep[e.g.,][]{Suwa2009,Liu2017b}. \citet{Liu2017b} calculated the dependence of the GW strains from NDAFs on both the BH spin and accretion rate. They demonstrated that GWs from NDAFs might be detected at a distance of $\sim$ 100 kpc/$\sim$1Mpc by the advanced LIGO/Einstein Telescope (ET) with a typical frequency of $\sim$10-100 Hz. They made a comparison of GWs from different central engines of GRBs: NDAFs, BZ mechanisms (no GW emission), and millisecond magnetars. GWs from the central engines of adjacent GRBs might be used to determine whether there is an NDAF, BZ jets, or a magnetar. Furthermore, \citet{van Putten2003} studied the GWs from a magnetized torus around a rapidly rotating BH. They pointed out that the configuration of the accretion torus itself might develop to the large non-axisymmetries. The torus converts $\sim 10 \%$  BH spin energy into the gravitational radiation through a finite number of the multipole mass moments and, to less degree, into MeV neutrinos and winds. As demonstrated in \citet{van Putten2019a}, they estimated total GW energy $E_{\rm{GW}}\simeq (3.5\pm 1)\%M_{\odot }c^{2}$ from BH spin-down after post-merger in GW170817. GWs from the suspended accretion are expected to be detected by LIGO-Virgo up to the distances of about 100 Mpc \citep[e.g.,][]{van Putten2019b}.

This paper is the second work in a series on the BH hyperaccretion in collapsars. In \citet[][hereafter Paper \uppercase\expandafter{\romannumeral 1}]{Wei2019}, we investigated the MeV neutrino emission from NDAFs in the collapsar scenarios. In the initial hundreds of seconds of the accretion process, the mass supply rate of the massive progenitor is generally higher than the ignition rate of NDAFs, but the jets are generally choked in the envelope. Thus, only neutrinos can be emitted from the center of a collapsar. We studied the effects of the masses and metallicities of the progenitor stars on the time-integrated spectra of electron neutrinos from NDAFs. The masses of collapsars have little influence on the neutrino spectrum, and the low metallicities are beneficial for the production of low-energy ($\lesssim$ 1 MeV) neutrinos. We also studied the differences in the electron neutrino spectra between NDAFs and proto-NSs (PNSs), which may help one verify the possible remnants of the core collapse of massive stars with future neutrino detectors.

In this paper, we focus on the GW emission from NDAFs in the collapsar scenarios and study the effects of the masses and metallicities of the progenitor stars on the GW emission from NDAFs. Meanwhile, a comparison of GW signals from NDAFs, jets, and collapsars is displayed.  The paper is organized as follows. In section 2, we describe the progenitor model and show the method to calculate the GWs emitted by anisotropic neutrino emission. Based on the time evolution of the mass accretion of progenitors with different masses and metallicities, the GW emission of NDAFs in the collapsars scenarios is studied. In addition, the detection of GW signals is discussed. In section 3, we compare the GW emission from different phases of collapsars. A summary is presented in section 4.

\section{GWs from NDAFs in collapsars}
\subsection{Progenitor model}

We adopt the pre-supernova (pre-SN) model \citep[see, e.g.,][]{Woosley2002,Woosley2007,Heger2010} in this work. After a massive star collapses, a rotating stellar-mass BH surrounded by a hyperaccretion disk might form. Using the density profiles of the pre-SN model (for details, see Paper \uppercase\expandafter{\romannumeral 1}), we can calculate the mass supply rate of the progenitors \citep[see, e.g.,][]{Suwa2011,Woosley2012,Matsumoto2015,Liu2018,Liu2019}, i.e.,
\beq
\dot{M}_{\rm{pro}}=\frac{dM_{r}}{dt_{\rm{ff}}}=\frac{dM_r/dr}{dt_{\rm{ff}}/dt}=\frac{2M_r}{t_{\rm{ff}}}(\frac{\rho}{\bar{\rho}-\rho }),
\eeq
where $M_r$ is the mass coordinate, $\rho$ is the mass density of the progenitor star, and $\bar{\rho}=3M_r/(4\pi r^{3})$ is the mean density within $r$. Here, we roughly set the accretion rate $\dot{M}$ equal to the mass supply rate \citep[e.g.,][]{Kashiyama2013,Nakauchi2013}. The accretion timescale of each mass shell at radius $r$ is roughly equal to the free-fall timescale:
\beq
t_{\rm{ff}}=\sqrt{\frac{3\pi}{32G\bar{\rho}}}=\frac{\pi}{2}\sqrt{\frac{r^{3}}{2GM_r}}.
\eeq

In the collapsar scenarios, the duration of the central engine can be related to the fallback accretion of a progenitor envelope. For the suspended accretion \citep{van Putten Ostriker2001}, the activity duration of the central engine is expected to depend on the lifetime of the BH spin. Such case is not considered in this work. In the initial hundreds of seconds of the accretion process, the jets are generally choked in the envelope of a collapsar, so no electromagnetic counterparts of the central engine can be observed \citep[see, e.g.,][]{Kashiyama2013,Nakauchi2013,Liu2018,Liu2019,Song2019}. Whether the jets can breakout out depends on the activity timescale of the central engine, the scale and density of the dense circumstellar medium (CSM) and the properties of the jets. If the activity of the central engine lasts long enough to allow the jets to break out of the envelope and CSM, an energetic LGRB will be triggered \citep[e.g.,][]{Liu2018,Liu2019}.

\subsection{GWs from NDAFs}

\begin{figure*}
\begin{minipage}{0.5\linewidth}
  \centerline{\includegraphics[angle=0,height=6cm,width=8cm]{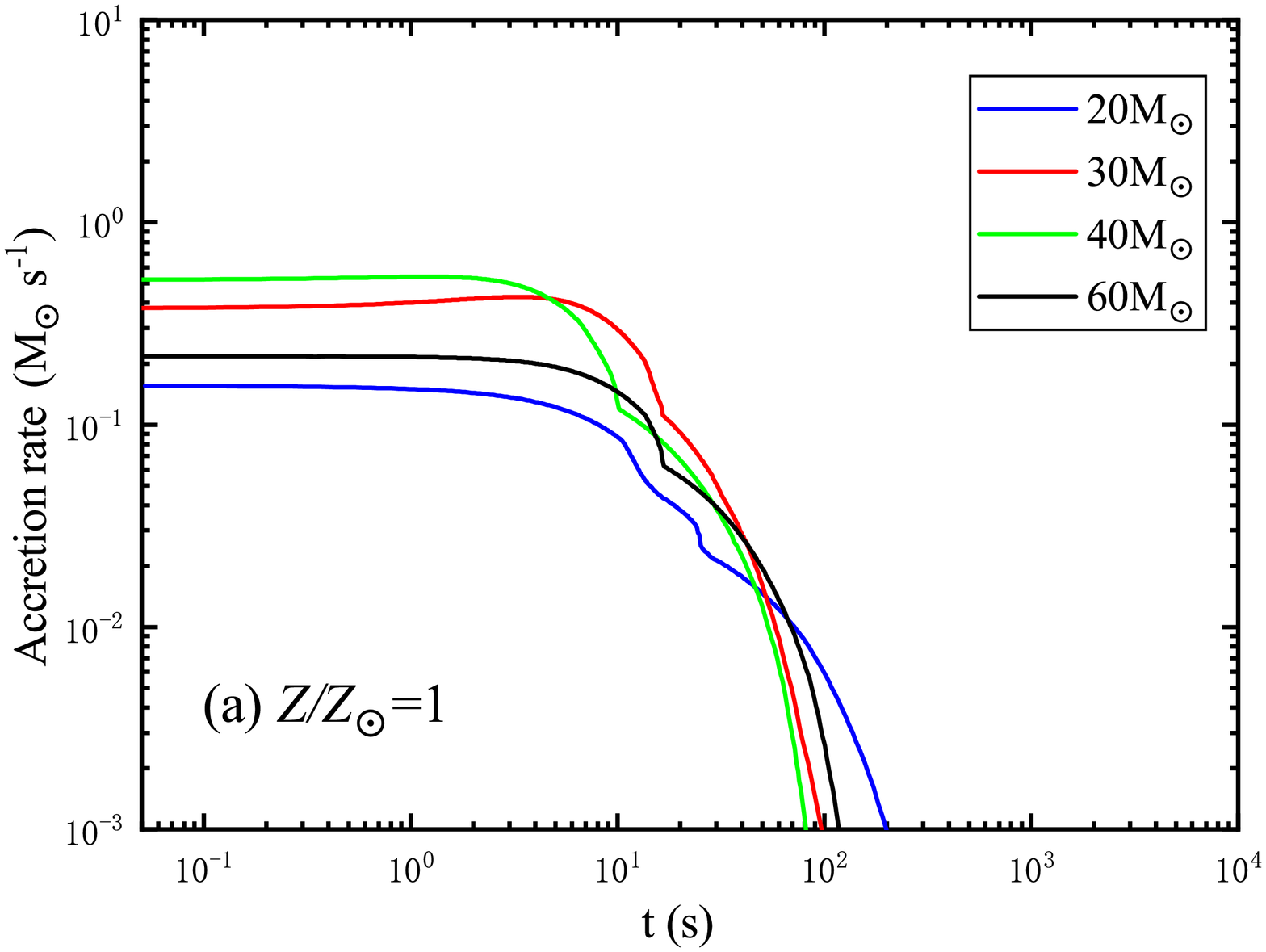}}
\end{minipage}
\hfill
\begin{minipage}{0.5\linewidth}
  \centerline{\includegraphics[angle=0,height=6cm,width=8cm]{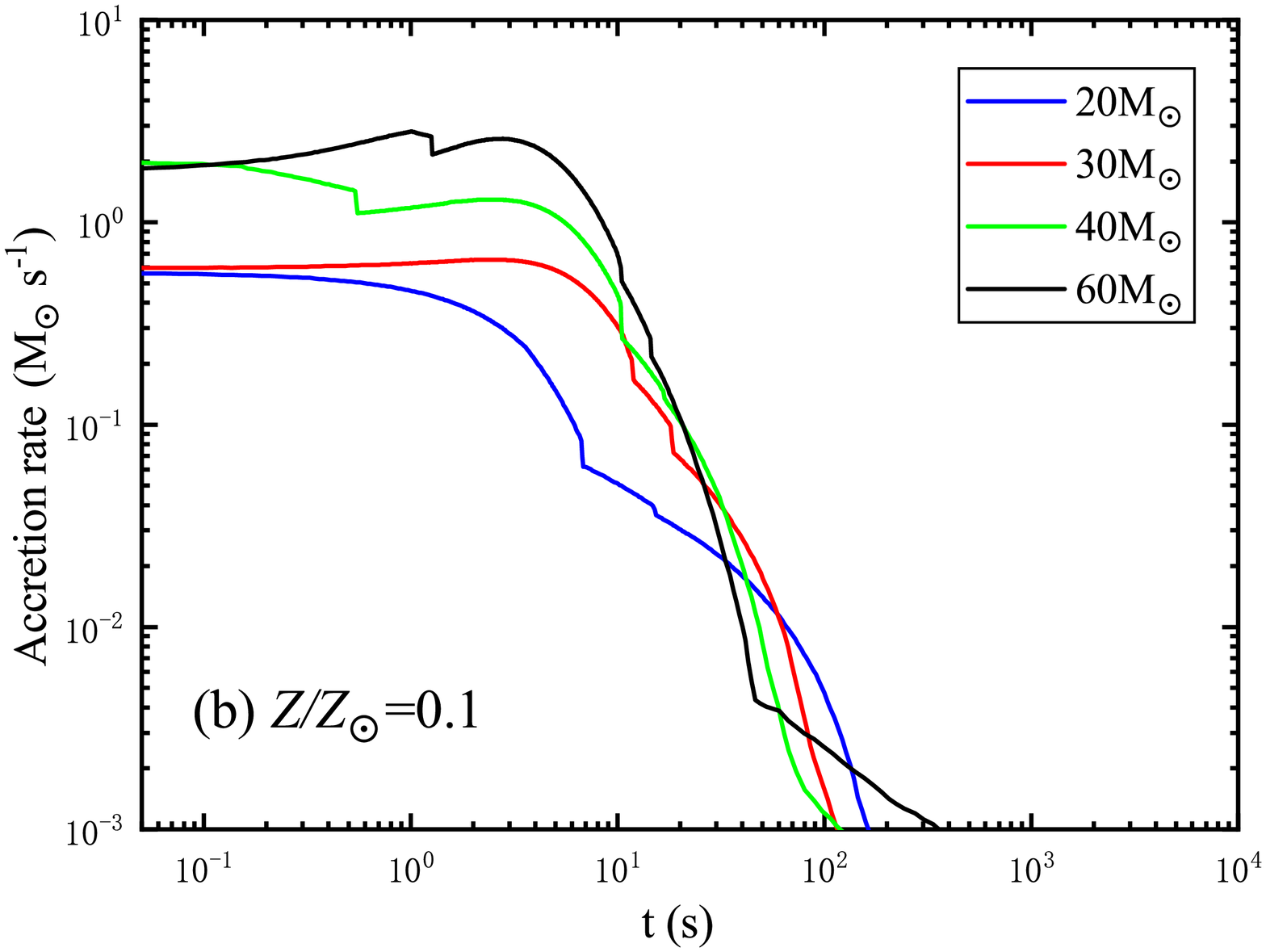}}
\end{minipage}
\vfill
\begin{minipage}{0.5\linewidth}
  \centerline{\includegraphics[angle=0,height=6cm,width=8cm]{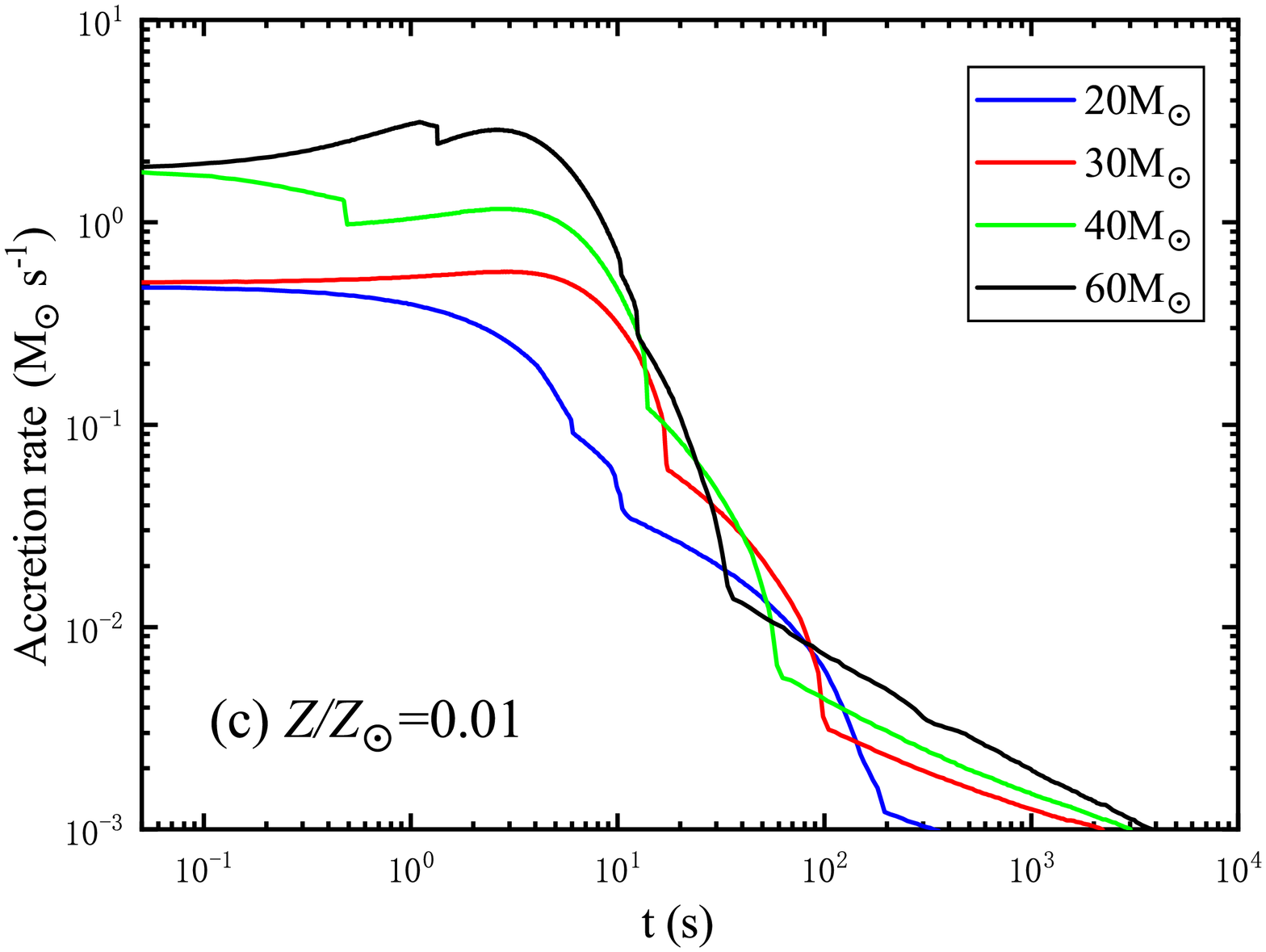}}
\end{minipage}
\hfill
\begin{minipage}{0.5\linewidth}
  \centerline{\includegraphics[angle=0,height=6cm,width=8cm]{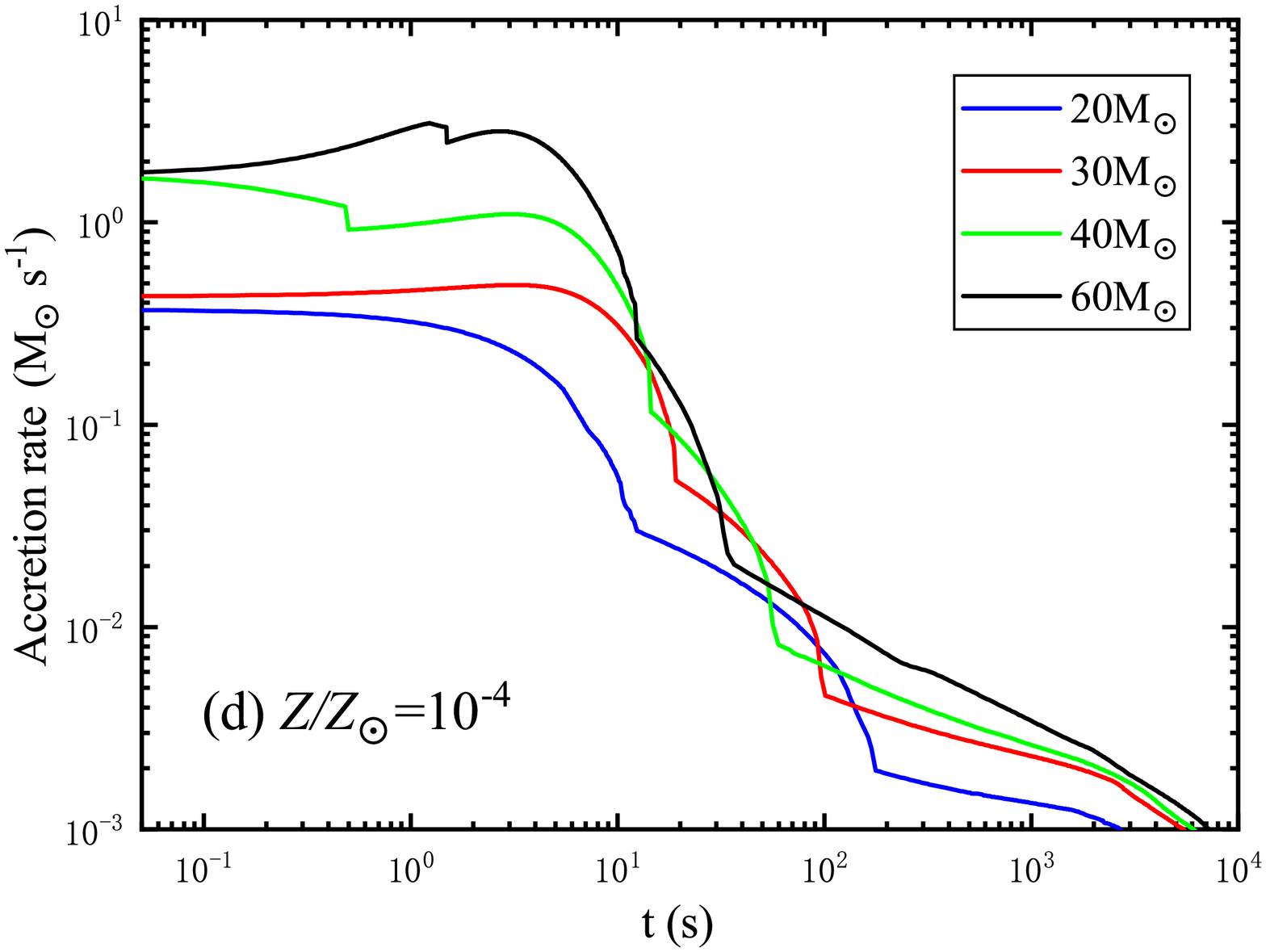}}
\end{minipage}
\caption{Time evolution of the mass accretion rate (mass supply rate) of progenitor stars with different masses $M_{\rm{pro}}$ and metallicities $Z$. The blue, red, green, and black curves correspond to progenitor masses of $M_{\rm{pro}}/M_{\odot}$=20, 30, 40, and 60, respectively.}
\end{figure*}

For NDAFs, there are some characteristic radii \citep[e.g.,][]{Chen2007,Zalamea2011,Liu2017,Liu2018,Zhang2018}, such as the ignition radius $r_{\rm{ign}}$, which can be defined as the radius such that $Q_{\nu}^{-}/Q_{\rm vis}=1/2$, where $Q_{\nu}^{-}$ and $Q_{\rm vis}$ are the neutrino cooling rate and the viscous heating rate, respectively \citep[e.g.,][]{Chen2007,Liu2017}. Inside the region of $\lesssim r_{\rm ign}$, one can say that neutrino cooling processes are dominant. Therefore, an NDAF is ignited only when $r_{\rm ign}$ is larger than the inner radius of the disk. The corresponding mass accretion rate is $\dot{M}_{\rm ign}$, which is mainly related to the BH spin and the viscous parameter of the disk. If $\dot{M}<\dot{M}_{\rm ign} \sim 0.001~ M_\odot~\rm s^{-1}$, the neutrino emission can be ignored, and the disk is no longer called an NDAF. In the collapsar scenarios, the mass accretion onto the BH decreases over time, and the typical duration of the NDAF in the collapsar is hundreds of seconds.

\citet{Xue2013} investigated one-dimensional global solutions of NDAFs in the Kerr metric, taking into account the detailed neutrino physics, chemical potential equilibrium, neutrino trapping and nucleosynthesis. They calculated 16 solutions with different characterized accretion rates and BH spins. Based on the results, they fitted time-independent analytical formulas, and the neutrino luminosity ${L}_{\rm{\nu }}$ is given by
\beq
\log {L}_{\rm{\nu }}({\rm erg\,s^{-1}})\approx 52.5+1.17a_{*}+1.17\log\,\dot{m},
\eeq
where $a_{*}$ ($0< a_{*}< 1$) is the mean dimensionless BH spin parameter, and $\dot{m}=\dot{M}/ M_{\odot }\rm ~s^{-1}$ is the dimensionless accretion rate. We adopt $a_{*}=0.9$ in our calculations.

According to the above formulas, we can roughly obtain the time evolution of the neutrino luminosity $L_{\rm{\nu}}(t)$. The GW emission from NDAFs depends on the neutrino luminosity, and the typical GW frequency is related to the variabilities and duration of neutrino emission.

The GWs from anisotropic neutrino radiation were first analyzed by \citet{Epstein1978}. We adopt the methods applied to CCSNe \citep[e.g.,][]{Burrows1996,Mueller1997,Kotake2006,Kotake2007} to calculate the GWs from NDAFs in the collapsar scenarios.

With the angles $\theta'$ and $\phi'$ defining the radiation direction in the source coordinate frame, the GW amplitude is given by \citep[e.g.,][]{Mueller1997,Suwa2009}
\beq
h_{+}(t,\vartheta)=&&\frac{2G}{Rc^{4}}\int_{-\infty}^{t-R/c }dt'\int_{4\pi }d\Omega '\Psi (\theta ',\phi ',\vartheta )\nonumber \\&&\times\frac{dL_{\rm{\nu}}(\theta ',t')}{d\Omega '},
\eeq
where $\vartheta$ is the viewing angle, $R$ is the distance from the observer to the source, $dL_{\rm{\nu}}/d\Omega '$ represents the direction-dependent neutrino luminosity per unit of solid angle in the direction of $\Omega'$, and $\Psi (\theta ',\phi ',\vartheta )$ denotes the angle dependent factor,
\beq
&&\Psi (\theta ',\phi ',\vartheta )=(1+\rm{cos}\theta '\rm{cos}\vartheta+\rm{sin}\theta '\rm{cos}\phi '\rm{sin}\vartheta ) \nonumber \\&& \times \frac{(\rm{sin}\theta '\rm{cos}\phi '\rm{cos}\vartheta -\rm{cos}\theta '\rm{sin}\vartheta )^{2}-\rm{sin}^{2}\theta '\rm{sin}^{2}\phi '}{(\rm{sin}\theta '\rm{cos}\phi '\rm{cos}\vartheta -\rm{cos}\theta '\rm{sin}\vartheta )^{2}+\rm{sin}^{2}\theta '\rm{sin}^{2}\phi '}.
\eeq

When $dL_{\rm{\nu}}/d\Omega '$ is axisymmetric, the counterpart of the amplitude, $h_{\times}^{TT}$, vanishes \citep[for details, see][]{Suwa2009}. In this work, we suppose the NDAF as a geometrically infinitely thin disk and assume that the emission of neutrinos is isotropic at any point of the disk surface. The neutrino luminosity per solid angle can be written as ${dL_{\rm{\nu }}}/{d\theta '}={L_{\rm{\nu }}}\left | \rm{cos} \theta '\right |/{2\pi }$. Then, Equation (4) is integrated analytically as
\beq
h_{+}(t,\vartheta)=&&\frac{1+2 \cos\vartheta }{3}\tan^{2}(\frac{\vartheta }{2})\frac{2G}{Rc^{4}} \nonumber \\&& \times \int_{-\infty }^{t-R/c}L_{\rm{\nu}}(t')dt'.
\eeq
Here, we can see the dependence of the GW amplitude on the viewing angle $\vartheta$. For $\vartheta=\pi/2$, the observer is located in the equatorial plane of the disk, and the GW amplitude is the largest.

The local energy flux of GWs is given by \citep[e.g.,][]{Suwa2009,Liu2017b}
\beq
\frac{dE_{\rm{GW}}}{R^{2}d\Omega dt}=\frac{c^{3}}{16\pi G}\left | \frac{d}{dt}h_{+}(t,\vartheta) \right |^{2},
\eeq
where $\Omega$ is the solid angle in the observer coordinate frame.

The total GW energy can be obtained by
\beq
E_{\rm{GW}}=\frac{\beta G}{9c^{5}}\int_{-\infty }^{\infty }dtL_{\rm{\nu}}(t)^{2},
\eeq
where $\beta \sim 0.47039$.

To obtain a GW spectrum, $L_{\rm{\nu}}(t)$ is written in terms of the inverse Fourier transform as
\beq
L_{\rm{\nu} }(t)=\int_{-\infty }^{+\infty} \tilde{L}_{\rm{\nu} }(f)e^{-2\pi ift}df;
\eeq
then, one can deduce the GW energy spectrum as
\beq
\frac{dE_{\rm{GW}}(f)}{df}=\frac{2\beta G}{9c^{5}}\left |  \tilde{L}_{\rm{\nu} }(f)\right |^{2}.
\eeq

To estimate the detectability of the GWs, the characteristic GW strains are defined by
\beq
h_{\rm{c}}(f)=\frac{1}{R}\sqrt{\frac{2}{\pi ^{2}}\frac{G}{c^{2}}\frac{dE_{\rm{GW}}(f)}{df}}
\eeq
for a given frequency $f$ \citep{Flanagan1998}. Since we obtain the characteristic GW strains, the signal-to-noise ratios (SNRs) obtained from the matched filtering for the GW detectors can be calculated. The SNR for an optimally oriented source is
\beq
{\rm{SNR^{2}}}=\int_{0}^{\infty}d(\ln{f})\frac{h_{\rm{c}}(\emph{f})^{2}}{h_{\rm{n}}(\emph{f})^{2}},
\eeq
where $h_{\rm{n}}f=\sqrt{5fS_{\rm{h}}(f)}$ is the noise amplitude and $S_{\rm{h}}(f)$ is the spectral density of the strain noise in the detector at frequency $f$.

\subsection{Results}

\begin{figure*}
\begin{minipage}{0.5\linewidth}
  \centerline{\includegraphics[angle=0,height=6cm,width=8cm]{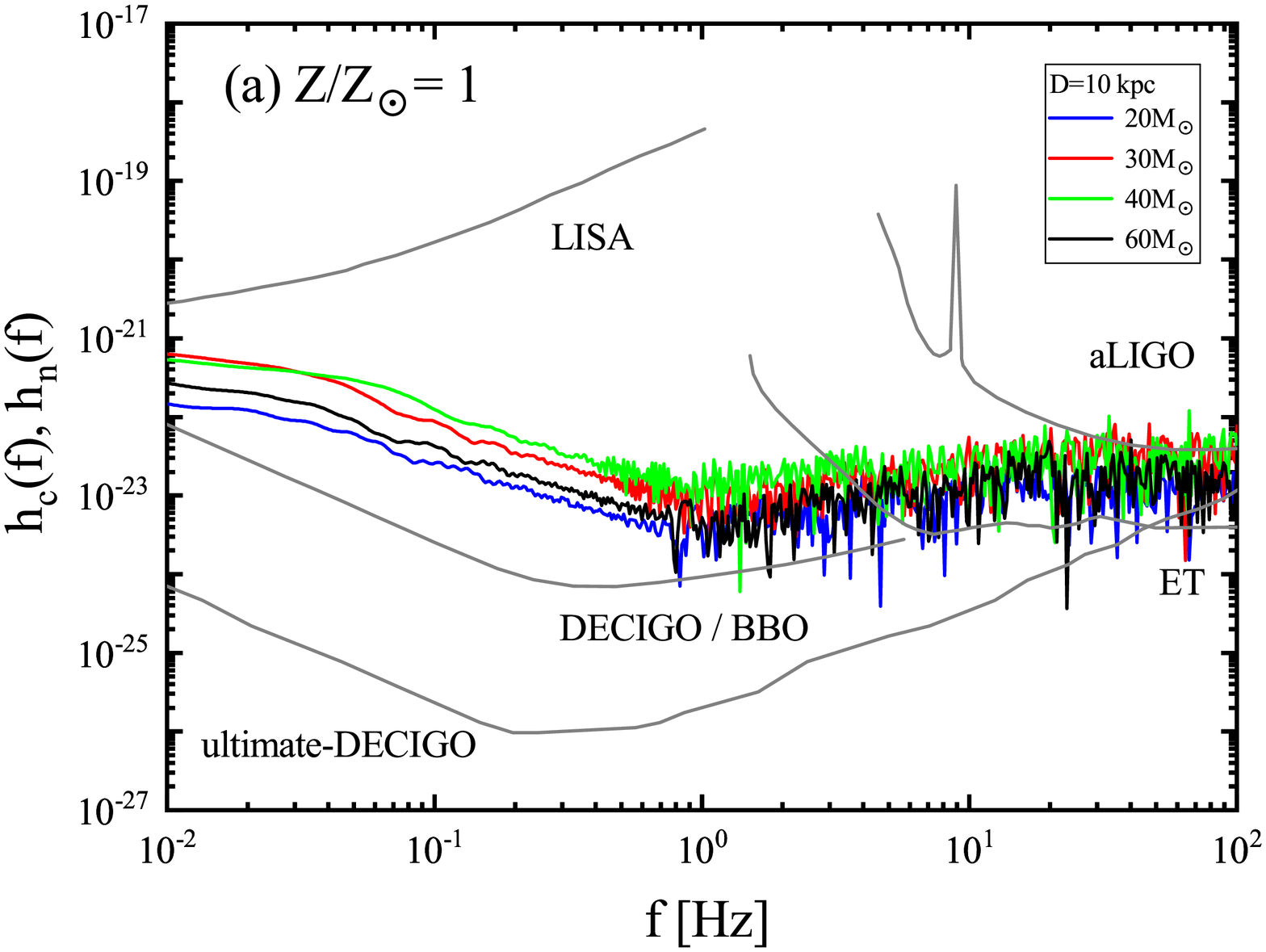}}
\end{minipage}
\hfill
\begin{minipage}{0.5\linewidth}
  \centerline{\includegraphics[angle=0,height=6cm,width=8cm]{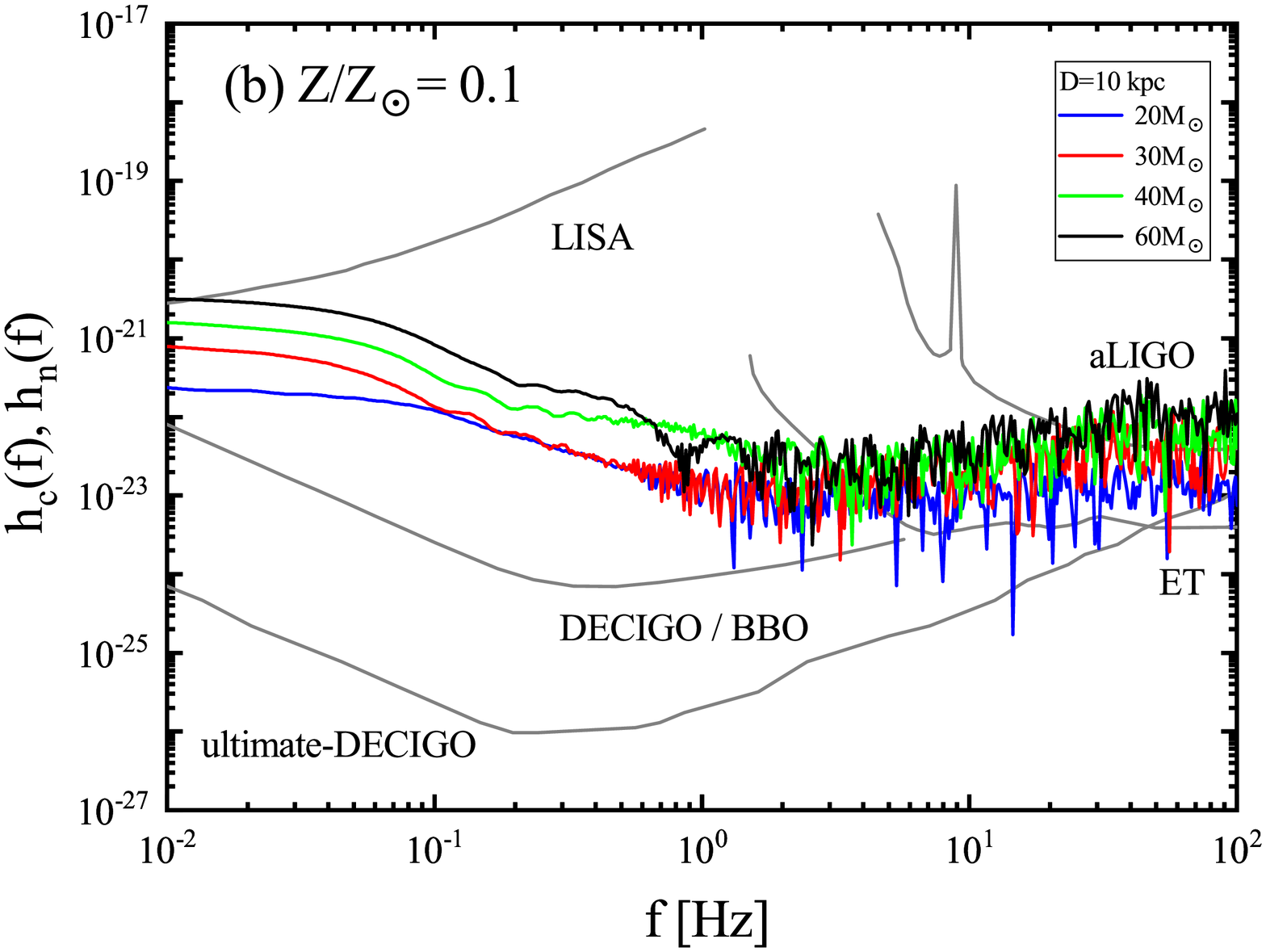}}
\end{minipage}
\vfill
\begin{minipage}{0.5\linewidth}
  \centerline{\includegraphics[angle=0,height=6cm,width=8cm]{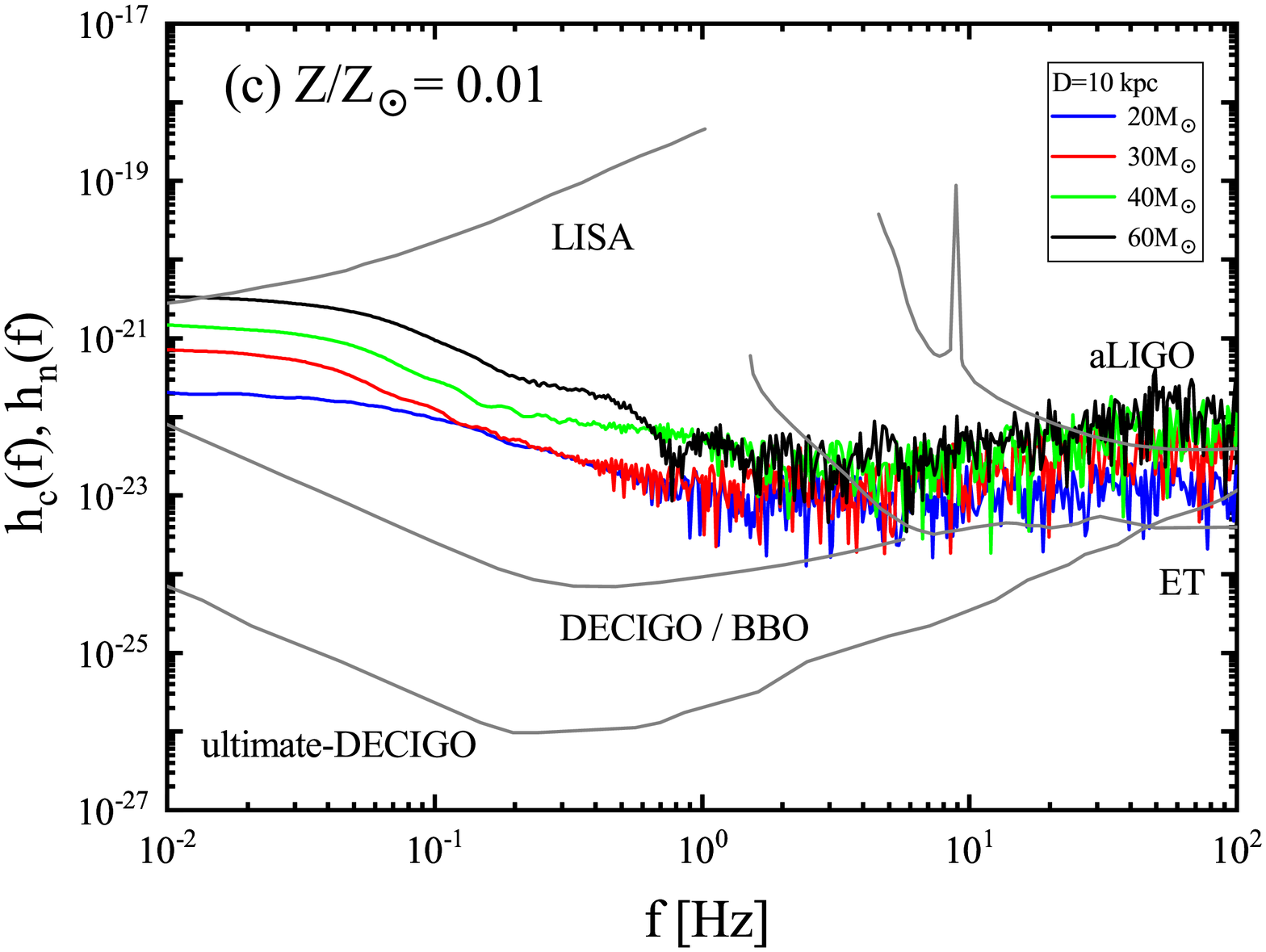}}
\end{minipage}
\hfill
\begin{minipage}{0.5\linewidth}
  \centerline{\includegraphics[angle=0,height=6cm,width=8cm]{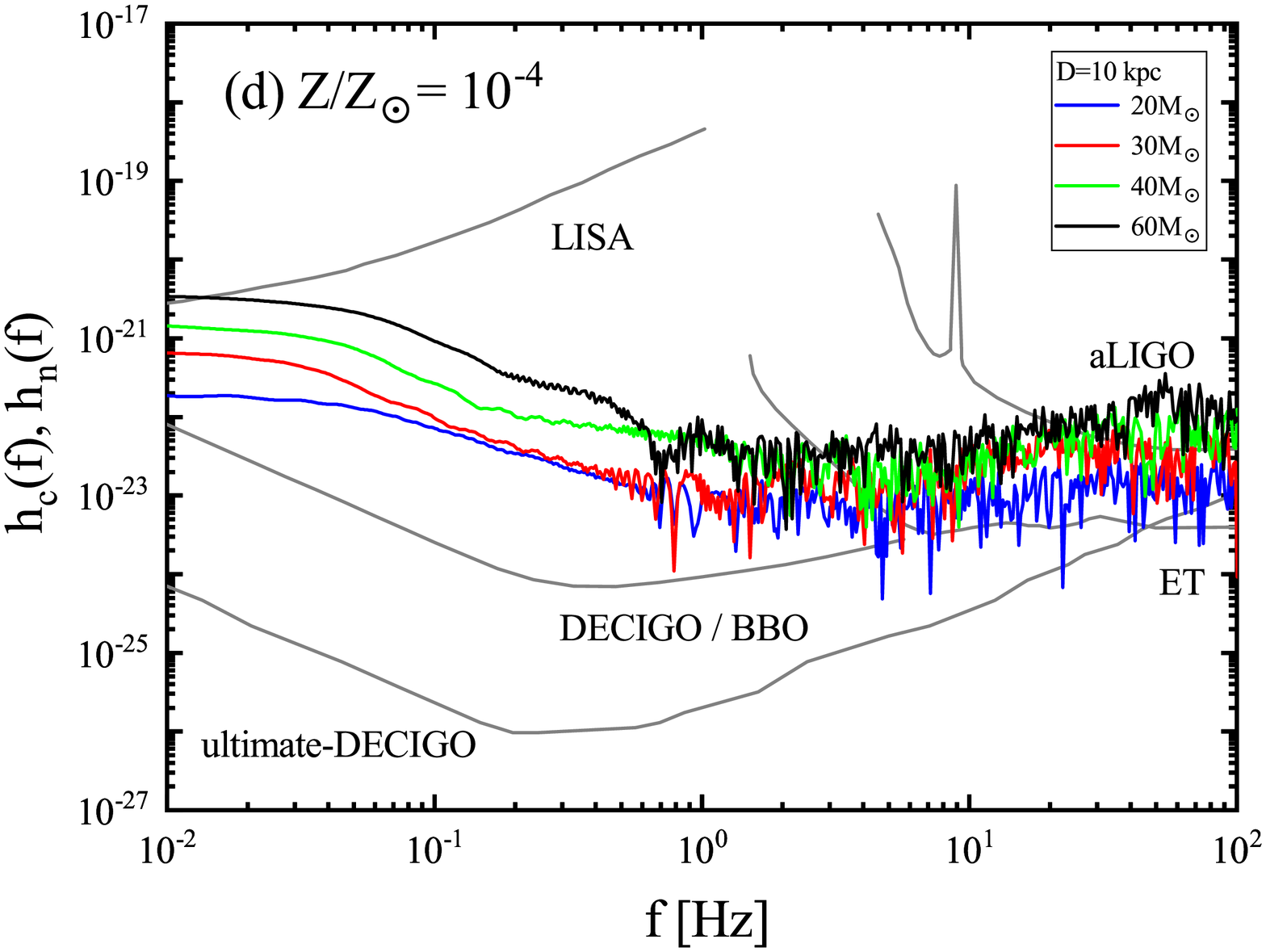}}
\end{minipage}
\caption{The strains of the GWs from NDAFs in the center of collapsars at the distance of 10 kpc. The blue, red, green, and black curves correspond to progenitor masses of $M_{\rm{pro}}/M_{\odot}$=20, 30, 40, and 60, respectively. In all four figures, the gray lines show the sensitivity lines (the noise amplitudes $h_{\rm{n}}$) of aLIGO, ET, LISA, DECIGO/BBO, and ultimate-DECIGO.}
\end{figure*}

We selected progenitor metallicities of $Z/Z_{\odot}$=1, 0.1, 0.01, and $10^{-4}$ and masses of $M_{\rm{pro}}/M_{\odot}$=20, 30, 40, and 60, where $Z_{\odot}$ is the metallicity of the Sun, to investigate the effects of the mass and metallicity on the GW emission of NDAFs. Based on the density profiles of the progenitor stars with different masses and metallicities (for details, see Paper \uppercase\expandafter{\romannumeral 1}), we can calculate the time evolution of the mass accretion rate of progenitor stars, as shown in Figure 1. The blue, red, green, and black curves correspond to progenitor masses of $M_{\rm{pro}}/M_{\odot}$=20, 30, 40, and 60, respectively. The different density profiles of the progenitor stars cause the difference in $\dot{M}$. In the initial accretion stage, the mass accretion rates are all approximately 1 $M_{\odot}~\rm{s}^{-1}$, which are typical mass accretion rates of NDAFs. The metallicities can affect the duration of the neutrino emission in collapsars, while the progenitor mass plays an important role in the time evolution of the mass accretion rate.

Figure 2 shows the strains of the GWs from NDAFs in the center of collapsars at the distance of 10 kpc. The blue, red, green, and black curves correspond to the progenitor masses of $M_{\rm{pro}}/M_{\odot}$=20, 30, 40, and 60, respectively. Sensitivity lines (the noise amplitudes $h_{\rm{n}}$) of aLIGO, ET, LISA, DECIGO/BBO, and ultimate-DECIGO are shown in these figures. As the progenitor mass increases, the mass accretion rate onto the BH tends to slightly rise in the initial accretion phase, as shown in Figure 1, so the GW strains increase within an order of magnitude. Meanwhile, the progenitor metallicities also have little influence on the GW strains, as the accretion rates in the initial accretion stage show little difference for the different metallicities and the same mass. Although the lower metallicities correspond to a longer duration of neutrino emission, the neutrino cooling is invalid in the late accretion stage. Overall, the GW signals from NDAFs in the center of the massive progenitor stars are more likely to be detected at the distance of 10 kpc.

The effects of the distance on the detection of GW signals from NDAFs are clearly displayed in Figure 3. At a distance of $\sim$ 10 kpc, the GWs from NDAFs in the center of the very massive progenitors might be detected by DECIGO/BBO and ultimate-DECIGO and by ET and aLIGO in the detectable frequency range of $\sim$ 10-100 Hz. Even so, it is difficult to distinguish the characteristics of the progenitor stars from the GW detection, as shown in Figure 2. Multimessenger observations, including electromagnetic and neutrino emissions, are indispensable for constraining the nature of the progenitor stars, as well as that of the central BH hyperaccretion systems. Moreover, one can see that the GW signals of NDAFs at the distance of 1 Mpc are almost impossible to detect by the operational or planned detectors.

Note that the star rotation was neglected in the above pre-SN model. For the moderately rotating stars, the rotation has limited effects on the neutrino emission from NDAFs. However, as shown in \citet{Janiuk2008}, the various rotation profiles imposed on the collapsing stars may significantly change the evolution of GRBs. Thus, for fast-rotating progenitor stars, the rotation might significantly change the GW signals from NDAFs.

\section{GW emission in massive collapsars}

GW emission from collapsars has been investigated for approximately four decades. Numerous numerical simulations predicted that GWs from various phenomena associated with gravitational collapse could be detected by ground-based and space-based interferometric GW detectors \citep[see the review by][]{Fryer2011}. For most of the massive stars, the GW signals from core collapsars will be similar to those from normal CCSNe, whose GW signals have been widely investigated \citep[see the review by][]{Ott2009,Kotake2013}. If the core collapsars and/or the resulting supernova (SN) explosions are nonspherical such that the third time derivative of the quadrupole moment of the mass-energy distribution is nonzero, part of the liberated gravitational binding energy will be emitted in the form of GWs. Such nonsphericities can be caused by the effects of rotation, convection and anisotropic neutrino emission. These effects lead either to small-scale statistical mass-energy fluctuations or large-scale asphericities \citep[e.g.,][]{Mueller1982,Moenchmeyer1991,Yamada1995,Zwerger1997,Rampp1998,Dimmelmeier2002,Fryer2002,Fryer2004b,Ott2004,Herant1995,Epstein1978,Burrows1996,Mueller1997,Muller2004}. In the collapsar scenarios, a massive star will go through the collapse, bounce, and postbounce phase, then BH formation, the hyperaccretion phase, and the GRB jet phase \citep[e.g.,][]{Kotake2012,Ott2011}. From the perspective of detecting GWs, we divide this evolutionary process into three periods: the collapsar phase (from collapse and bounce to BH formation), central engine phase (hyperaccretion phase) and GRB jet phase. We study the typical frequencies and amplitudes of GW signals from these three different phases and plot the results in Figure 4. The blue, purple, and orange shaded regions represent the collapsar phase, NDAFs and GRB jet phase, respectively. The detection distance is 10 kpc. The characteristic amplitude of GWs from the suspend accretion is not include is Figure 4. The GWs from the non-axisymmetric accretion flow around a rapidly rotating BH are expected to detected up to distance of about 100 Mpc \citep{van Putten2019b}. At the distance of 10 kpc, the characteristic amplitude of the GWs from the suspend accretion is much larger than that from NDAFs.

\begin{figure}
\includegraphics[angle=0,height=6.5cm,width=8.5cm]{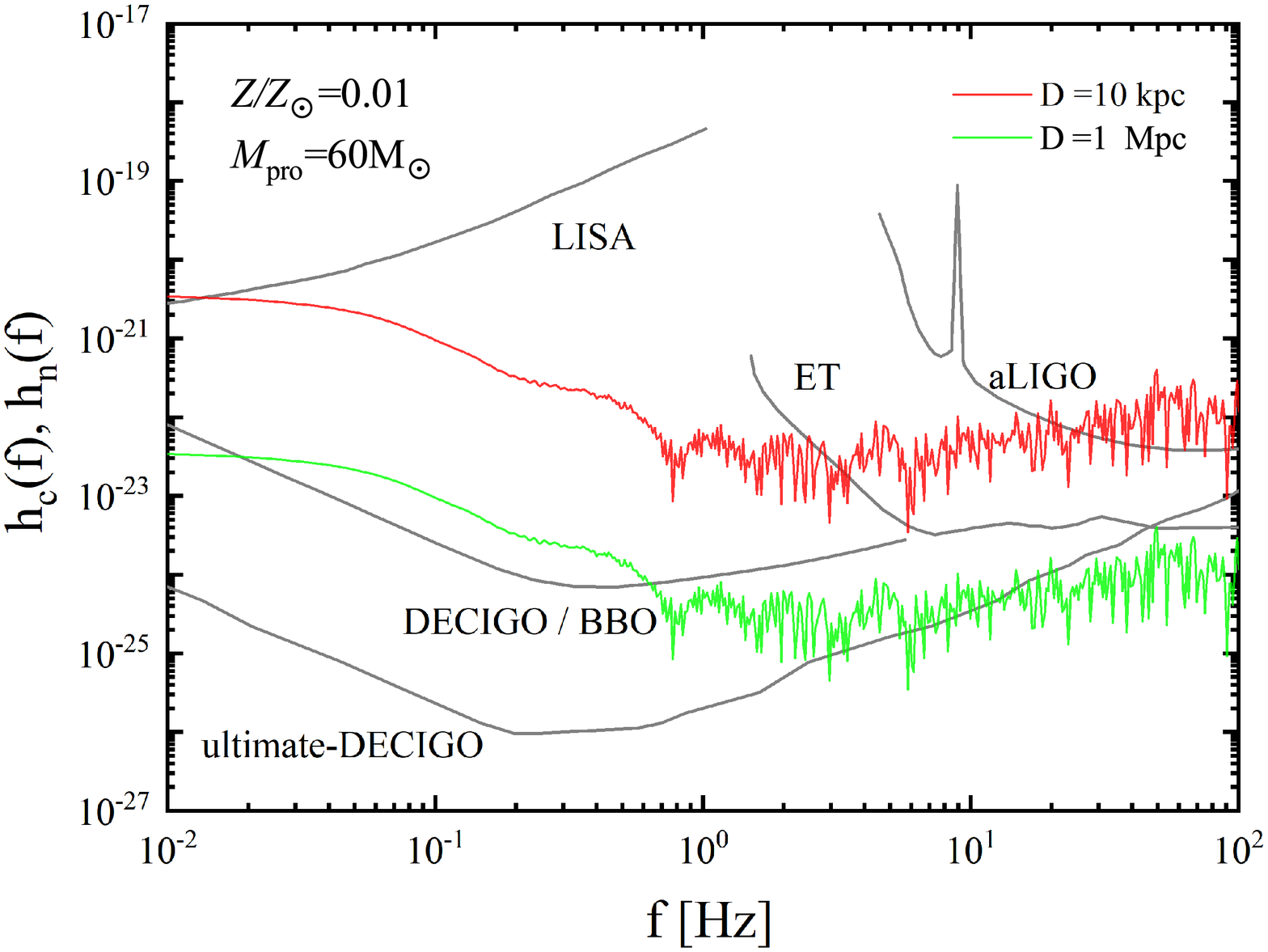}
\caption{The dependence of the GW strains on the distances. The mass and metallicity of the progenitors are $M_{\rm{pro}}/M_{\odot}$=60 and $Z/Z_{\odot}$= 0.01, respectively.}
\end{figure}

In the collapsar phase, the primary GW signals are unlikely to be different from normal CCSNe. Most of the original studies paid attention to the strong GW signals produced at the rotating collapse and core bounce phase due to the large-scale aspherical motion of matter. The peak amplitude is roughly proportional to the collapsar spin. The typical frequency is expected to be 100-1000 Hz \citep[see, e.g.,][]{Kotake2006,Kotake2013,Ott2009}.

In the postbounce phase, the anisotropic matter motions associated with the convection, anisotropic neutrino radiation, and standing-accretion-shock instability \citep[SASI, e.g.,][]{Scheck2008,Burrows2006,Foglizzo2007,Blondin2003,Blondin2006,Kotake2007} are expected to be the primary sources for GWs. Convective instability is an important feature of the postbounce evolution of massive stars \citep[see, e.g.,][]{Janka2007,Buras2006,Burrows2006,Burrows2007,Janka1996}. The convective overturn is expected as the entropy- and lepton-gradient-driven prompt convection that may occur immediately after bounce, lepton-gradient driven PNS convection, and neutrino-driven convection in the post-shock heating region \citep[e.g.,][]{Ott2008,Fryer2000,Burrows1983}. The SASI caused by either an advective acoustic or a purely acoustic feedback cycle would lead to the growth of perturbations in the stalled shock \citep[e.g.,][]{Scheck2008,Foglizzo2007,Blondin2003,Blondin2006}. When the SASI enters a nonlinear phase, it would heavily distort the post-shock region and affect convection. Both the convection and SASI would lead to time-varying mass-quadrupole moments giving rise to a long-duration stage of large amplitude GW emission. A semiquantitative summary of the GW emission by the aspherical fluid motions associated with the convection and SASI is given by \citet{Ott2009}. Based on numerous previous simulations \citep[e.g.,][]{Ott2006,Muller2004,Marek2009}, they provided estimations for the typical GW strains at 10 kpc, with the typical emission frequency $f$ approximately in the range of 100-1000 Hz.

For core collapse of massive stars, anisotropic neutrino emission may arise (a) from the global asymmetries in the (precollapse) matter distribution \citep[see, e.g.,][]{Burrows1996,Fryer2004a}, (b) from the convective overturn and SASI  \citep[see, e.g.,][]{Ott2006,Kotake2007,Marek2009}, and (c) from the rotationally deformed PNSs \citep[e.g.,][]{Kotake2006,Muller2004}. In contrast to the rapidly varying GW waveforms from matter motion, the GW waveforms from the anisotropic neutrino emission show a long-timescale variability. Hence, the anisotropic neutrino emission dominates the GW spectrum at low frequencies (below $\sim$ 100 Hz) \citep[e.g.,][]{Burrows1996,Muller2004,Kotake2006}. In addition, the precollapse density inhomogeneities  \citep[e.g.,][]{Burrows1996,Mueller1997,Fryer2004}, nonaxisymmetric rotational instabilities \citep[e.g.,][]{Rampp1998,Ott2007}, g-mode \citep[e.g.,][]{Ott2006} and r-mode pulsations of PNSs \citep[e.g.,][]{Andersson2011}, and aspherical mass ejection may contribute to the overall GW signature.

In the later BH formation phase, the typical frequency of GW signals is relatively high \citep[e.g.,][]{Pan2018}. GW emission at BH formation in the collapsar scenario has been studied by some previous works \citep[e.g.,][]{Sekiguchi2005,Ott2011,Cerd2013,Pan2018}. These works showed that the peak frequency from g-mode PNS oscillations at BH formation is expected to be above 2 kHz. However, such GW signals are very close to the limit of the current GW detectors and are difficult to detect. Overall, according to the GW emission mechanisms and current GW detectors, the most promising detectable frequency is at 100-1000 Hz in the collapsar phase. For a rotational core-collapse event, the average maximum amplitude of GWs at the distance $R$ is calculated as \citep{Dimmelmeier2002,Moore2015}
\beq
h_{\rm{max}}=8.9\times 10^{-21}(\frac{10 \rm{kpc}}{R}).
\eeq
As shown in Figure 4, the GW signals from collapsars are likely to be detected by aLIGO and ET.

\begin{figure}
\centering
\includegraphics[angle=0,scale=0.35]{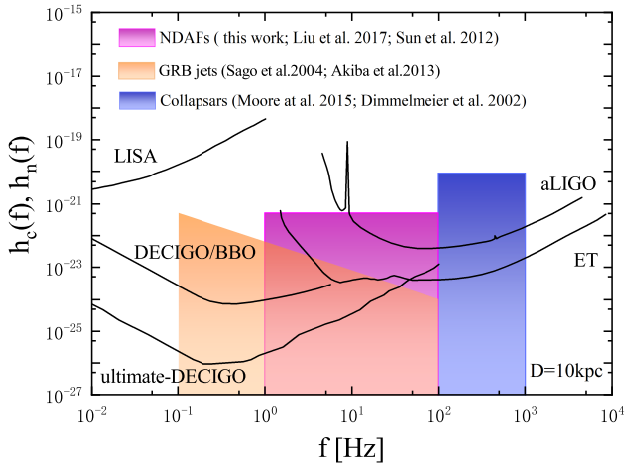}
\caption{The characteristic amplitude of GWs from different sources in a collapsar. The blue, purple, and orange shaded boxes represent the collapsar phase, NDAFs and GRB jet phase, respectively. The gray lines show the sensitivity lines (the noise amplitudes $h_{\rm n}$) of aLIGO, ET, LISA, DECIGO/BBO, and ultimate-DECIGO. The distance is 10 kpc.}
\end{figure}

In the central engine phase (hyperaccretion phase), GWs are expected to be from the BH-inner-disk precession system \citep[e.g.,][]{Sun2012} and the anisotropic neutrino emission from NDAFs \citep[e.g.,][]{Suwa2009,Liu2017}. In \citet{Sun2012}, they studied the GWs of the jet precession based on NDAFs around spinning BHs. They argued that disk-driven jet precession may be common in a BH accretion system since the only necessary condition is that the angular momentum of the initial accretion flow is misaligned with the BH spinning axis. The GW signals from such systems are expected to be detectable at the frequency of tens of Hz and have comparable amplitudes to GW signals from the anisotropic neutrino emission. For GWs from the anisotropic neutrino emission, the detectable frequency is at 1-100 Hz, as shown in Figure 2. As discussed in section 2.3, the progenitor mass and metallicity have little influence on GW signals. The maximum amplitude of GWs from NDAFs is roughly $h_{\rm{max}}=5\times 10^{-22}$.

In the GRB jet phase, the relativistic jets are expected to be GW sources and have been studied by some previous works \citep[see, e.g.,][]{Segalis2001,Sago2004,Birnholtz2013,Akiba2013}. \citet{Sago2004} analyzed the GWs from the accelerating phase of GRB jets based on the internal shock model. The ultrarelativistic nonspherically symmetrical acceleration of energetic jets is expected to emit GWs. For such GW signals, the maximum amplitude is $\sim 10^{-22}$ at the frequency of $\sim$ 0.1 Hz at 10 kpc. Meanwhile, GW emission is also expected to be produced in the decelerating phase of GRB jets \citep[e.g.,][]{Akiba2013}. The GW amplitude is approximately $\sim 10^{-24}$ Hz at the frequency of 10-100 Hz, which is too low to be detected. Therefore, in such a phase, the GW signals are more likely to be detected at 0.1-10 Hz by DECIGO/BBO and ultimate-DECIGO.

As the above discussion indicates, the GW signals related to the various mechanisms from the three phases have different characteristic frequencies. The collapsar phase occurs earlier than the central engine phase and GRB jet phase. From the perspective of detection, one would receive the high-frequency GW signals from collapsars first and then the low-frequency GW signals from later phases.

\section{Summary}

In this work, we employed the pre-SN model and studied the GWs generated by the anisotropic neutrino emission from NDAFs in the center of collapsars. We found that the progenitor mass and metallicity have little influence on the GW emission. The GW signals from NDAFs in the center of the massive progenitor stars are more likely to be detected by GW detectors at the distance of 10 kpc. Then, we briefly summarized the GW emission in the different phases of collapsars. The primary detectable frequencies and strains in the three phases (the collapsar, central engine, and GRB jet phases) are different. Considering that the three phases occur in a time sequence, one may distinguish the detectable GWs from the different phases, which can partly verify the collapsar model and BH hyperaccretion solution.

Furthermore, it is inadequate to constrain the nature of the progenitors solely according to the GW detection. By combining the electromagnetic counterparts, neutrinos, and GWs, we might obtain the accurate and authentic properties of the progenitor stars and central BH accretion systems. In \citet{Song2019}, they constrained the characteristics of the progenitor stars of the GRB-SN case by LGRB-SN data. In Paper \uppercase\expandafter{\romannumeral 1}, we have investigated the effects of the mass and metallicity of progenitor stars on the time-integrated spectrum of electron neutrinos from NDAFs. The detection of sub-MeV neutrinos may help us limit the metallicities of the progenitor stars.

\acknowledgments
We appreciate the assistance of Prof. A. Heger who provided the pre-SN data. This work was supported by the National Natural Science Foundation of China under grant 11822304 and the Fundamental Research Funds for the Central Universities at Xiamen University under grant 20720190115.


\begin{thebibliography}{99}
\bibitem[Abbott et al.(2016)]{Abbott2016} Abbott, B.~P., Abbott, R., Abbott, T.~D., et al.\ 2016, \prl, 116, 061102
\bibitem[Abbott et al.(2017)]{Abbott2017} Abbott, B.~P., Abbott, R., Abbott, T.~D., et al.\ 2017, \prl, 119, 161101
\bibitem[Akiba et al.(2013)]{Akiba2013} Akiba, S., Nakada, M., Yamaguchi, C., et al.\ 2013, \pasj, 65, 59
\bibitem[Andersson et al.(2011)]{Andersson2011} Andersson, N., Ferrari, V., Jones, D.~I., et al.\ 2011, General Relativity and Gravitation, 43, 409
\bibitem[Barkov \& Komissarov(2008)]{Barkov2008} Barkov, M.~V., \& Komissarov, S.~S.\ 2008, \mnras, 385, L28
\bibitem[Birnholtz \& Piran(2013)]{Birnholtz2013} Birnholtz, O., \& Piran, T.\ 2013, \prd, 87, 123007
\bibitem[Blandford \& Znajek(1977)]{Blandford1977} Blandford, R.~D., \& Znajek, R.~L.\ 1977, \mnras, 179, 433
\bibitem[Blondin \& Mezzacappa(2006)]{Blondin2006} Blondin, J.~M., \& Mezzacappa, A.\ 2006, \apj, 642, 401
\bibitem[Blondin et al.(2003)]{Blondin2003} Blondin, J.~M., Mezzacappa, A., \& DeMarino, C.\ 2003, \apj, 584, 971
\bibitem[Buras et al.(2006)]{Buras2006} Buras, R., Janka, H.-T., Rampp, M., et al.\ 2006, \aap, 457, 281
\bibitem[Burrows \& Hayes(1996)]{Burrows1996} Burrows, A., \& Hayes, J.\ 1996, \prl, 76, 352
\bibitem[Burrows \& Lattimer(1983)]{Burrows1983} Burrows, A., \& Lattimer, J.~M.\ 1983, \apj, 270, 735
\bibitem[Burrows et al.(2006)]{Burrows2006} Burrows, A., Livne, E., Dessart, L., et al.\ 2006, \apj, 640, 878
\bibitem[Burrows et al.(2007)]{Burrows2007} Burrows, A., Livne, E., Dessart, L., et al.\ 2007, \apj, 655, 416
\bibitem[Cerd{\'a}-Dur{\'a}n et al.(2013)]{Cerd2013} Cerd{\'a}-Dur{\'a}n, P., DeBrye, N., Aloy, M.~A., et al.\ 2013, \apjl, 779, L18
\bibitem[Chen \& Beloborodov(2007)]{Chen2007} Chen, W.-X., \& Beloborodov, A.~M.\ 2007, \apj, 657, 383
\bibitem[Cutler \& Thorne(2002)]{Cutler2002} Cutler, C., \& Thorne, K.~S.\ 2002, General Relativity and Gravitation, 72
\bibitem[Dimmelmeier et al.(2002)]{Dimmelmeier2002} Dimmelmeier, H., Font, J.~A., \& M{\"u}ller, E.\ 2002, \aap, 393, 523
\bibitem[Epstein(1978)]{Epstein1978} Epstein, R.\ 1978, \apj, 223, 1037
\bibitem[Flanagan \& Hughes(1998)]{Flanagan1998} Flanagan, {\'E}. {\'E}., \& Hughes, S.~A.\ 1998, \prd, 57, 4535
\bibitem[Foglizzo et al.(2007)]{Foglizzo2007} Foglizzo, T., Galletti, P., Scheck, L., et al.\ 2007, \apj, 654, 1006
\bibitem[Fruchter et al.(2006)]{Fruchter2006} Fruchter, A.~S., Levan, A.~J., Strolger, L., et al.\ 2006, \nat, 441, 463
\bibitem[Fryer(2004)]{Fryer2004} Fryer, C.~L.\ 2004, \apjl, 601, L175
\bibitem[Fryer \& Heger(2000)]{Fryer2000} Fryer, C.~L., \& Heger, A.\ 2000, \apj, 541, 1033
\bibitem[Fryer et al.(2002)]{Fryer2002} Fryer, C.~L., Holz, D.~E., \& Hughes, S.~A.\ 2002, \apj, 565, 430
\bibitem[Fryer et al.(2004a)]{Fryer2004a} Fryer, C.~L., Holz, D.~E., \& Hughes, S.~A.\ 2004a, \apj, 609, 288
\bibitem[Fryer et al.(2004b)]{Fryer2004b} Fryer, C.~L., Holz, D.~E., Hughes, S.~A., et al.\ 2004b, Astrophysics and Space Science Library, 373
\bibitem[Fryer \& New(2011)]{Fryer2011} Fryer, C.~L., \& New, K.~C.~B.\ 2011, Living Reviews in Relativity, 14, 1
\bibitem[Galama et al.(1998)]{Galama1998} Galama, T.~J., Vreeswijk, P.~M., van Paradijs, J., et al.\ 1998, \nat, 395, 670
\bibitem[Gu et al.(2006)]{Gu2006} Gu, W.-M., Liu, T., \& Lu, J.-F.\ 2006, \apjl, 643, L87
\bibitem[Heger \& Woosley(2010)]{Heger2010} Heger, A., \& Woosley, S.~E.\ 2010, \apj, 724, 341
\bibitem[Herant(1995)]{Herant1995} Herant, M.\ 1995, \physrep, 256, 117
\bibitem[Hjorth et al.(2003)]{Hjorth2003} Hjorth, J., Sollerman, J., M{\o}ller, P., et al.\ 2003, \nat, 423, 847
\bibitem[Janiuk \& Proga(2008)]{Janiuk2008} Janiuk, A., \& Proga, D.\ 2008, \apj, 675, 519
\bibitem[Janiuk et al.(2007)]{Janiuk2007} Janiuk, A., Yuan, Y., Perna, R., et al.\ 2007, \apj, 664, 1011
\bibitem[Janka et al.(2007)]{Janka2007} Janka, H.-T., Langanke, K., Marek, A., et al.\ 2007, \physrep, 442, 38
\bibitem[Janka \& M{\"u}ller(1996)]{Janka1996} Janka, H.-T., \& M{\"u}ller, E.\ 1996, \aap, 306, 167
\bibitem[Kashiyama et al.(2013)]{Kashiyama2013} Kashiyama, K., Nakauchi, D., Suwa, Y., et al.\ 2013, \apj, 770, 8
\bibitem[Kawanaka \& Mineshige(2007)]{Kawanaka2007} Kawanaka, N., \& Mineshige, S.\ 2007, \apj, 662, 1156
\bibitem[Kohri \& Mineshige(2002)]{Kohri2002} Kohri, K., \& Mineshige, S.\ 2002, \apj, 577, 311
\bibitem[Kotake(2013)]{Kotake2013} Kotake, K.\ 2013, Comptes Rendus Physique, 14, 318
\bibitem[Kotake et al.(2007)]{Kotake2007} Kotake, K., Ohnishi, N., \& Yamada, S.\ 2007, \apj, 655, 406
\bibitem[Kotake et al.(2006)]{Kotake2006} Kotake, K., Sato, K., \& Takahashi, K.\ 2006, Reports on Progress in Physics, 69, 971
\bibitem[Kotake et al.(2012)]{Kotake2012} Kotake, K., Takiwaki, T., \& Harikae, S.\ 2012, \apj, 755, 84
\bibitem[Kumar \& Zhang(2015)]{Kumar2015} Kumar, P., \& Zhang, B.\ 2015, \physrep, 561, 1
\bibitem[Lee et al.(2000a)]{Lee2000a} Lee, H.~K., Brown, G.~E., \& Wijers, R.~A.~M.~J.\ 2000a, \apj, 536, 416
\bibitem[Lee et al.(2000b)]{Lee2000b} Lee, H.~K., Wijers, R.~A.~M.~J., \& Brown, G.~E.\ 2000b, \physrep, 325, 83
\bibitem[Lee et al.(2005)]{Lee2005} Lee, W.~H., Ramirez-Ruiz, E., \& Page, D.\ 2005, \apj, 632, 421
\bibitem[Lei et al.(2009)]{Lei2009} Lei, W.~H., Wang, D.~X., Zhang, L., et al.\ 2009, \apj, 700, 1970
\bibitem[Lei et al.(2013)]{Lei2013} Lei, W.-H., Zhang, B., \& Liang, E.-W.\ 2013, \apj, 765, 125
\bibitem[Lei et al.(2017)]{Lei2017} Lei, W.-H., Zhang, B., Wu, X.-F., et al.\ 2017, \apj, 849, 47
\bibitem[Liu et al.(2007)]{Liu2007} Liu, T., Gu, W.-M., Xue, L., et al.\ 2007, \apj, 661, 1025
\bibitem[Liu et al.(2017a)]{Liu2017} Liu, T., Gu, W.-M., \& Zhang, B.\ 2017a, \nar, 79, 1
\bibitem[Liu et al.(2017b)]{Liu2017b} Liu, T., Lin, C.-Y., Song, C.-Y., et al.\ 2017b, \apj, 850, 30
\bibitem[Liu et al.(2015)]{Liu2015} Liu, T., Lin, Y.-Q., Hou, S.-J., et al.\ 2015, \apj, 806, 58
\bibitem[Liu et al.(2019)]{Liu2019} Liu, T., Song, C.-Y., Yi, T., et al.\ 2019, Journal of High Energy Astrophysics, 22, 5
\bibitem[Liu et al.(2018)]{Liu2018} Liu, T., Song, C.-Y., Zhang, B., et al.\ 2018, \apj, 852, 20
\bibitem[Liu et al.(2016)]{Liu2016} Liu, T., Zhang, B., Li, Y., et al.\ 2016, \prd, 93, 123004
\bibitem[MacFadyen \& Woosley(1999)]{MacFadyen1999} MacFadyen, A.~I., \& Woosley, S.~E.\ 1999, \apj, 524, 262
\bibitem[Malesani et al.(2004)]{Malesani2004} Malesani, D., Tagliaferri, G., Chincarini, G., et al.\ 2004, \apjl, 609, L5
\bibitem[Marek et al.(2009)]{Marek2009} Marek, A., Janka, H.-T., \& M{\"u}ller, E.\ 2009, \aap, 496, 475
\bibitem[Matsumoto et al.(2015)]{Matsumoto2015} Matsumoto, T., Nakauchi, D., Ioka, K., et al.\ 2015, \apj, 810, 64
\bibitem[Maurer et al.(2010)]{Maurer2010} Maurer, J.~I., Mazzali, P.~A., Deng, J., et al.\ 2010, \mnras, 402, 161
\bibitem[McKinney \& Gammie(2004)]{McKinney2004} McKinney, J.~C., \& Gammie, C.~F.\ 2004, \apj, 611, 977
\bibitem[Mizuno et al.(2004)]{Mizuno2004} Mizuno, Y., Yamada, S., Koide, S., et al.\ 2004, \apj, 615, 389
\bibitem[Modjaz et al.(2006)]{Modjaz2006} Modjaz, M., Stanek, K.~Z., Garnavich, P.~M., et al.\ 2006, \apjl, 645, L21
\bibitem[Moenchmeyer et al.(1991)]{Moenchmeyer1991} Moenchmeyer, R., Schaefer, G., M{\"u}ller, E., et al.\ 1991, \aap, 246, 417
\bibitem[Moore et al.(2015)]{Moore2015} Moore, C.~J., Cole, R.~H., \& Berry, C.~P.~L.\ 2015, Classical and Quantum Gravity, 32, 015014
\bibitem[M{\"u}ller(1982)]{Mueller1982} M{\"u}ller, E.\ 1982, \aap, 114, 53
\bibitem[M{\"u}ller \& Janka(1997)]{Mueller1997} M{\"u}ller, E., \& Janka, H.-T.\ 1997, \aap, 317, 140
\bibitem[M{\"u}ller et al.(2004)]{Muller2004} M{\"u}ller, E., Rampp, M., Buras, R., et al.\ 2004, \apj, 603, 221
\bibitem[Nagataki(2009)]{Nagataki2009} Nagataki, S.\ 2009, \apj, 704, 937
\bibitem[Nagataki(2018)]{Nagataki2018} Nagataki, S.\ 2018, Reports on Progress in Physics, 81, 026901
\bibitem[Nakauchi et al.(2013)]{Nakauchi2013} Nakauchi, D., Kashiyama, K., Suwa, Y., et al.\ 2013, \apj, 778, 67
\bibitem[Narayan et al.(2001)]{Narayan2001} Narayan, R., Piran, T., \& Kumar, P.\ 2001, \apj, 557, 949
\bibitem[O'Dea(2002)]{O'Dea2002} O'Dea, C.~P.\ 2002, \nar, 46, 41
\bibitem[Ott(2009)]{Ott2009} Ott, C.~D.\ 2009, Classical and Quantum Gravity, 26, 063001
\bibitem[Ott et al.(2006)]{Ott2006} Ott, C.~D., Burrows, A., Dessart, L., et al.\ 2006, \prl, 96, 201102
\bibitem[Ott et al.(2008)]{Ott2008} Ott, C.~D., Burrows, A., Dessart, L., et al.\ 2008, \apj, 685, 1069
\bibitem[Ott et al.(2004)]{Ott2004} Ott, C.~D., Burrows, A., Livne, E., et al.\ 2004, \apj, 600, 834
\bibitem[Ott et al.(2007)]{Ott2007} Ott, C.~D., Dimmelmeier, H., Marek, A., et al.\ 2007, \prl, 98, 261101
\bibitem[Ott et al.(2011)]{Ott2011} Ott, C.~D., Reisswig, C., Schnetter, E., et al.\ 2011, \prl, 106, 161103
\bibitem[Pan et al.(2018)]{Pan2018} Pan, K.-C., Liebend{\"o}rfer, M., Couch, S.~M., et al.\ 2018, \apj, 857, 13
\bibitem[Pian et al.(2006)]{Pian2006} Pian, E., Mazzali, P.~A., Masetti, N., et al.\ 2006, \nat, 442, 1011
\bibitem[Popham et al.(1999)]{Popham1999} Popham, R., Woosley, S.~E., \& Fryer, C.\ 1999, \apj, 518, 356
\bibitem[Rampp et al.(1998)]{Rampp1998} Rampp, M., Mueller, E., \& Ruffert, M.\ 1998, \aap, 332, 969
\bibitem[Sago et al.(2004)]{Sago2004} Sago, N., Ioka, K., Nakamura, T., et al.\ 2004, \prd, 70, 104012
\bibitem[Scheck et al.(2008)]{Scheck2008} Scheck, L., Janka, H.-T., Foglizzo, T., et al.\ 2008, \aap, 477, 931
\bibitem[Segalis \& Ori(2001)]{Segalis2001} Segalis, E.~B., \& Ori, A.\ 2001, \prd, 64, 064018
\bibitem[Sekiguchi \& Shibata(2005)]{Sekiguchi2005} Sekiguchi, Y.-I., \& Shibata, M.\ 2005, \prd, 71, 084013
\bibitem[Song \& Liu(2019)]{Song2019} Song, C.-Y., \& Liu, T.\ 2019, \apj, 871, 117
\bibitem[Song et al.(2016)]{Song2016} Song, C.-Y., Liu, T., Gu, W.-M., et al.\ 2016, \mnras, 458, 1921
\bibitem[Stanek et al.(2003)]{Stanek2003} Stanek, K.~Z., Matheson, T., Garnavich, P.~M., et al.\ 2003, \apjl, 591, L17
\bibitem[Sun et al.(2012)]{Sun2012} Sun, M.-Y., Liu, T., Gu, W.-M., et al.\ 2012, \apj, 752, 31
\bibitem[Suwa \& Ioka(2011)]{Suwa2011} Suwa, Y., \& Ioka, K.\ 2011, \apj, 726, 107
\bibitem[Suwa \& Murase(2009)]{Suwa2009} Suwa, Y., \& Murase, K.\ 2009, \prd, 80, 123008
\bibitem[van Putten(2001)]{van Putten2001} van Putten, M.~H.~P.~M.\ 2001, \physrep, 345, 1
\bibitem[van Putten et al.(2011)]{van Putten2011} van Putten, M.~H.~P.~M., Della Valle, M., \& Levinson, A.\ 2011, \aap, 535, L6
\bibitem[van Putten et al.(2019a)]{van Putten2019a} van Putten, M.~H.~P.~M., Della Valle, M., \& Levinson, A.\ 2019, \apjl, 876, L2
\bibitem[van Putten \& Levinson(2003)]{van Putten2003} van Putten, M.~H.~P.~M., \& Levinson, A.\ 2003, \apj, 584, 937
\bibitem[van Putten et al.(2019b)]{van Putten2019b} van Putten, M.~H.~P.~M., Levinson, A., Frontera, F., et al.\ 2019, European Physical Journal Plus, 134, 537
\bibitem[van Putten \& Ostriker(2001)]{van Putten Ostriker2001} van Putten, M.~H.~P.~M., \& Ostriker, E.~C.\ 2001, \apjl, 552, L31
\bibitem[Wei et al.(2019)]{Wei2019} Wei, Y.-F., Liu, T., \& Song, C.-Y.\ 2019, \apj, 878, 142
\bibitem[Woosley(1993)]{Woosley1993} Woosley, S.~E.\ 1993, \apj, 405, 273
\bibitem[Woosley \& Bloom(2006)]{Woosley2006} Woosley, S.~E., \& Bloom, J.~S.\ 2006, \araa, 44, 507
\bibitem[Woosley \& Heger(2007)]{Woosley2007} Woosley, S.~E., \& Heger, A.\ 2007, \physrep, 442, 269
\bibitem[Woosley \& Heger(2012)]{Woosley2012} Woosley, S.~E., \& Heger, A.\ 2012, \apj, 752, 32
\bibitem[Woosley et al.(2002)]{Woosley2002} Woosley, S.~E., Heger, A., \& Weaver, T.~A.\ 2002, Reviews of Modern Physics, 74, 1015
\bibitem[Wu et al.(2013)]{Wu2013} Wu, X.-F., Hou, S.-J., \& Lei, W.-H.\ 2013, \apjl, 767, L36
\bibitem[Xue et al.(2013)]{Xue2013} Xue, L., Liu, T., Gu, W.-M., et al.\ 2013, \apjs, 207, 23
\bibitem[Yamada \& Sato(1995)]{Yamada1995} Yamada, S., \& Sato, K.\ 1995, \apj, 450, 245
\bibitem[Zalamea \& Beloborodov(2011)]{Zalamea2011} Zalamea, I., \& Beloborodov, A.~M.\ 2011, \mnras, 410, 2302
\bibitem[Zhang(2018)]{Zhang2018} Zhang, B.\ 2018, The Physics of Gamma-Ray Bursts (Cambridge: Cambridge Univ. Press)
\bibitem[Zhang et al.(2004)]{Zhang2004} Zhang, W., Woosley, S.~E., \& Heger, A.\ 2004, \apj, 608, 365
\bibitem[Zwerger \& M{\"u}ller(1997)]{Zwerger1997} Zwerger, T., \& M{\"u}ller, E.\ 1997, \aap, 320, 209
\end{thebibliography}
\end{document}